\documentclass[aps,prb,preprint,groupedaddress,showpacs]{revtex4}
\usepackage{amsmath}
\usepackage{amsfonts}
\usepackage{amssymb}
\usepackage{graphicx}

\begin{document}


\title{Photoluminescence modification by a high-order photonic band with abnormal dispersion in ZnO inverse opal}


\author{Heeso Noh, Michael Scharrer, Mark A. Anderson, Robert P.H. Chang, Hui Cao}
\affiliation{Materials Research Institute, Northwestern University, Evanston, IL 60208}

\begin{abstract}
We measured the angle- and polarization-resolved reflection and photoluminescence spectra of ZnO inverse opals. Significant enhancement of spontaneous emission is observed. The enhanced emission not only has good directionality but also can be linearly polarized. A detailed theoretical analysis and numerical simulation reveal that such enhancement is caused by the abnormal dispersion of a high-order photonic band. The frozen mode at a stationary inflection point of a dispersion curve can strongly modify the intensity, directionality and polarization of spontaneous emission.  
\end{abstract}

\pacs{42.70.Qs, 42.25.Fx, 78.55.-m, 78.40.-q}

\maketitle

\section{Introduction}

Photonic crystals (PhCs) have been extensively studied for control of spontaneous emission. The first proposal by Yablonovitch in 1987 \cite{yablonovitch_inhibited_1987} utilized a complete photonic band gap (PBG) in a three-dimensional (3D) PhC, which was realized in 2004 by Noda and coworkers.\cite{ogawa_control_2004} Compared to one-dimensional and two-dimensional PhCs, the 3D PhC has the utmost control of light-matter interactions. However, it is much more difficult to fabricate 3D PhCs, especially with the top-down approach. An alternative approach based on self-assembly attracts much attention due to ease of fabrication. Such a bottom-up approach has been widely used for fabrication of face-centered-cubic (FCC) structures such as opals or inverse opals. An opal that is made of dielectric spheres has only partial PBG, namely, the PBG exists only within certain angle range instead of all angles. An inverse opal can have a full PBG if the refractive index contrast is high enough. It has been shown  
\cite{romanov_emission_2001,garcia_quantum_2006,koenderink_experimental_2003,bechger_directional_2005,yoshino_novel_1999,yamasaki_spontaneous_1998,romanov_effect_1997,pallavidino_synthesis_2006,lopez_photonic_1999,bogomolov_photonic_1997,lodahl_controlling_2004,wang_decay_2002,barth_spectral_2005,megens_light_1999,yoshino_observation_1998,nikolaev_quantitative_2005,koenderink_broadband_2002,wang_decay_2003,blanco_cds_1998,schriemer_modified_2000} 
that a partial PBG can effect the spontaneous emission process in an opal or inverse opal by suppressing spontaneous emission into certain directions. Enhancement of spontaneous emission is also observed at the edge of PBG. Most of these studies are focused on the lowest-order PBG in an opal or inverse opal. There are only a few experimental studies of the effect of high-order photonic bands on spontaneous emission. 
\cite{bechger_directional_2005,king_photoluminescence_2006,yang_enhanced_2006}
Recent studies 
\cite{baryshev_photonic_2004,rybin_complex_2006,galisteo-lopez_optical_2003,pavarini_band_2005,tarhan_photonic_1996,miguez_band_2004,ohtaka_photonic_1996,romanov_diffraction_2001,garcia_optical_2006,van_driel_multiple_2000,galisteo-lopez_high-energy_2004,galisteo-lopez_optical_2007,vos_higher_2000,balestreri_optical_2006,schroden_optical_2002,garcia-santamaria_optical_2005}
show that a high-order photonic band can have abnormal dispersion which leads to many interesting phenomena such as super prism
 \cite{kosaka_superprism_1998,ochiai_superprism_2001,prasad_superprism_2003}
and  negative refractive index
\cite{notomi_theory_2000,ren_three-dimensional_2007}. 
In this paper, we demonstrate that a high-order photonic band with abnormal dispersion can significantly modify the photoluminescence (PL) intensity, directionality and polarization in an inverse opal.       

The dispersion of photons in a PhC can be dramatically different from that in free space. Let us denote the dispersion of a photonic mode by $\omega(k)$, where $\omega$ is the photon frequency and $k$ is the wave vector. A mode with $d \omega / d k \simeq 0$ is called a slow mode because the group velocity is nearly zero. There have been many proposals of utilizing the slow modes of PhC to reduce the speed of light by orders of magnitude. A serious problem that hinders the slow light application is that a typical slow mode with $d^2 \omega / d k^2 \neq 0$ has large   impedance mismatch at the PhC/air interface, thus the conversion efficiency of incident light into the slow mode is very low. To solve this problem, Figotin and Vitebskiy proposed to use the photonic mode at the stationary inflection point of a dispersion curve of a photonic band.
\cite{figotin_oblique_2003,figotin_frozen_2006} Such a mode has both $d \omega / d k \simeq 0$ and $d^2 \omega / d k^2 \simeq 0$. It is called a frozen mode. When the incident light is in resonance with a frozen mode, the vanishing group velocity is offset by the diverging electromagnetic energy density. The energy flux inside the PhC is finite and comparable to the incident flux. Hence, the incident light can be completely converted to the frozen mode instead of being reflected. Our aim is to employ the unique properties of a frozen mode in a 3D PhC to tailor spontaneous emission. The vanishing group velocity enhances emission into the frozen mode, while the perfect impedance match at the PhC/air interface leads to efficient extraction of emission from PhC. 

The paper is organized as follows. Section II starts with a brief description of  sample fabrication process and experimental setup, followed by the experimental results of angle- and polarization-resolved reflection and PL spectra. In Section III, we calculate the reduced photonic band structure, density of states, and reflectivity to illustrate the effects of frozen modes on spontaneous emission process. To show the generality of such effects, we report PL enhancement in a different crystal direction in Section IV. Section V is the conclusion.   


\section{Experiment}

\subsection{Sample fabrication}

\begin{figure}[htbp]
\begin{tabular}{c c}
\includegraphics[width=80mm]{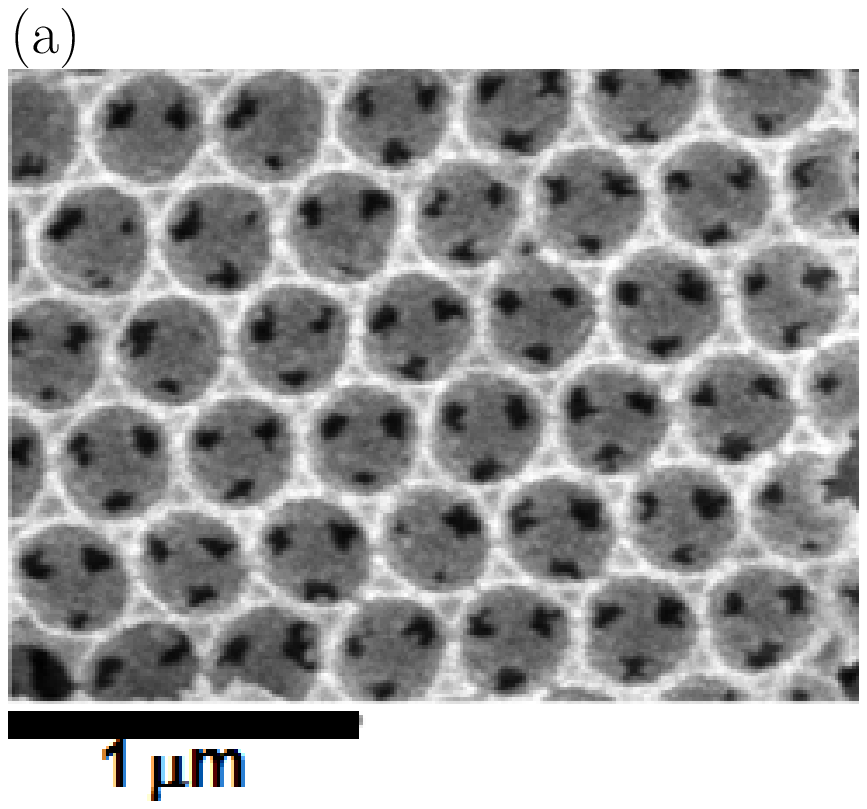} &
\includegraphics[width=80mm]{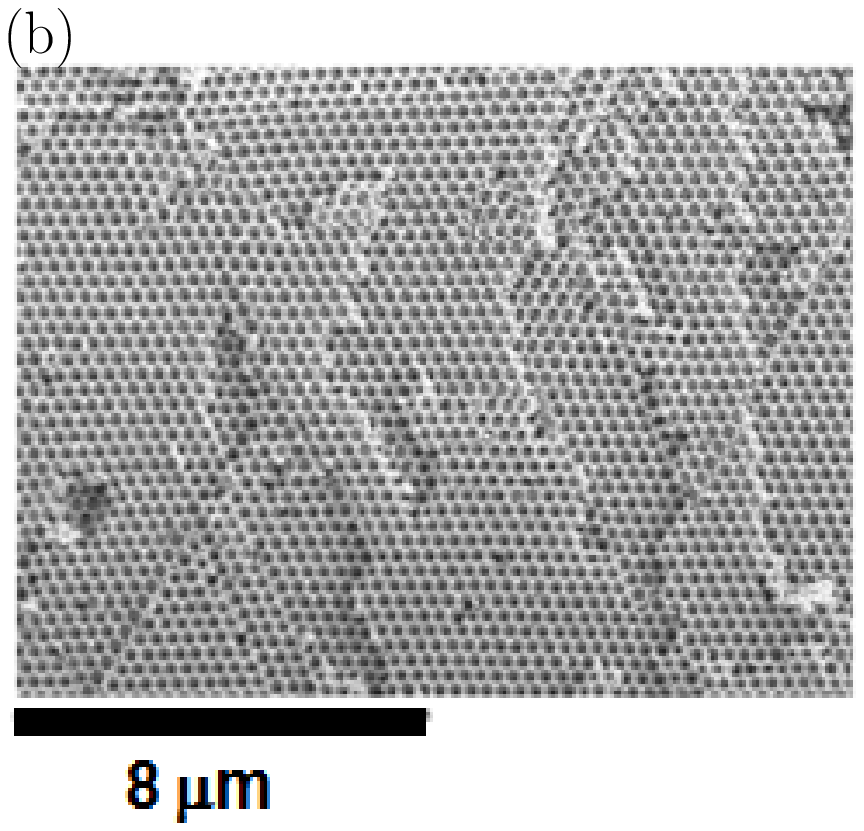} 
\end{tabular}
\caption{SEM images of the top surface (a) and cleaved edge (b) of a ZnO inverse opal.}
\label{fig:SEM}
\end{figure}

The fabrication of ZnO inverse opals is detailed in Refs.\cite{scharrer_fabrication_2005,scharrer_thesis_2007}. Monodisperse polystyrene spheres were assembled to face-centered-cubic (FCC) structure on a glass substrate via a vertical deposition process. The sample thickness could be varied from 20 to 100 layers of spheres. The opal had ``domains'' with widths of $\sim$ 50 $\mu$m and lengths of hundreds of micron, separated by cracks. The crystalline arrangement was constant across these cracks which  formed after the self-assembly process during drying of the opal. ZnO was infiltrated into the template by atomic layer deposition (ALD). We ensured the exposure times were sufficiently long during ALD growth so that the precursors could fully diffused into the opal structure and the ZnO thin film grew conformally and uniformly around each polystyrene sphere throughout the sample. The polystyrene spheres were then removed by firing at elevated temperature. Figure~\ref{fig:SEM} shows the scanning electron microscope (SEM) images of the top surface and cleaved edge of a ZnO inverse opal. The sample surface was parallel to (111) crystallographic plane. Even for a 100-layer-thick sample, ZnO was fully infiltrated into the opal template and the filling was nearly 100\%. 

A random sample was fabricated to provide reference for optical measurements. Polystyrene spheres of different sizes were mixed and deposited on a glass substrate. The lack of monodispersity prevented the formation of ordered crystal domains. The randomly packed structure was subsequently infiltrated with ZnO and fired at the same temperature as for ZnO inverse opal. The identical fabrication conditions ensured similar microstructure and material properties.

\subsection{Experimental setup}

\begin{figure}[htbp]
\begin{tabular}{c c}
\includegraphics[width=80mm]{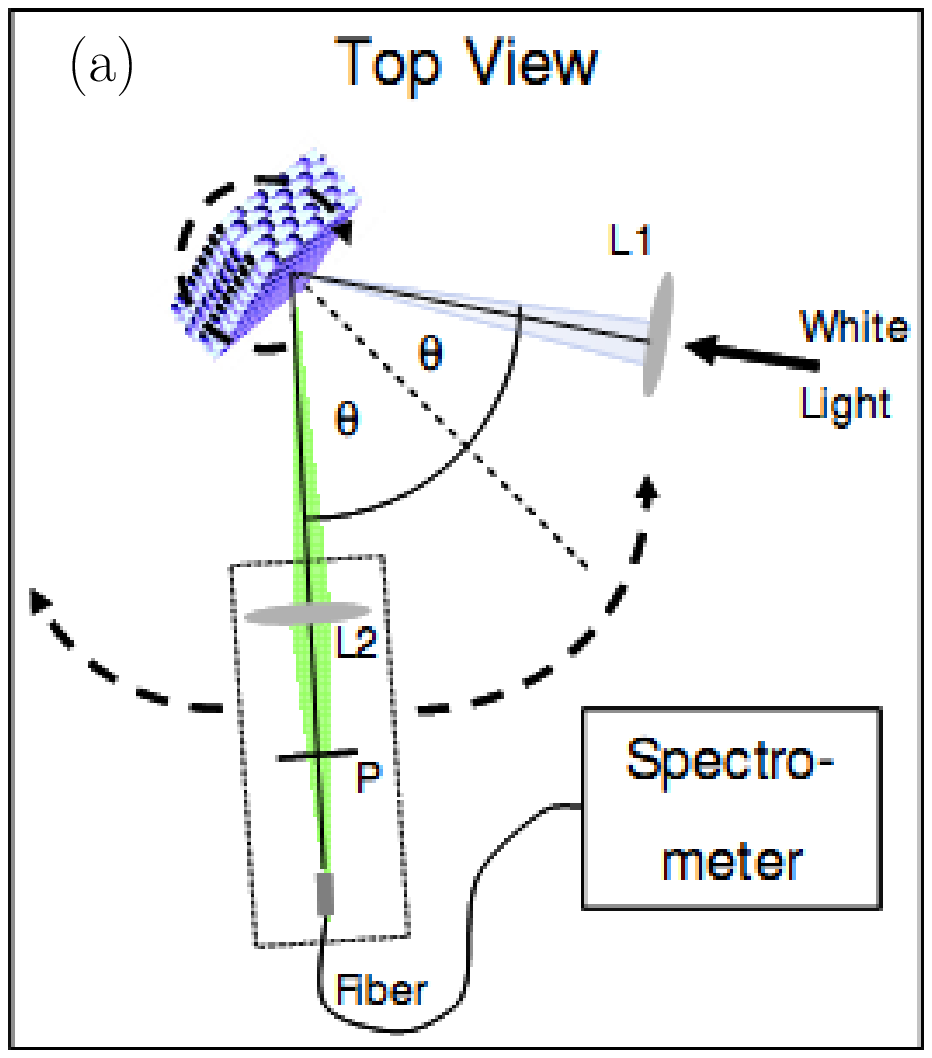} &
\includegraphics[width=80.5mm]{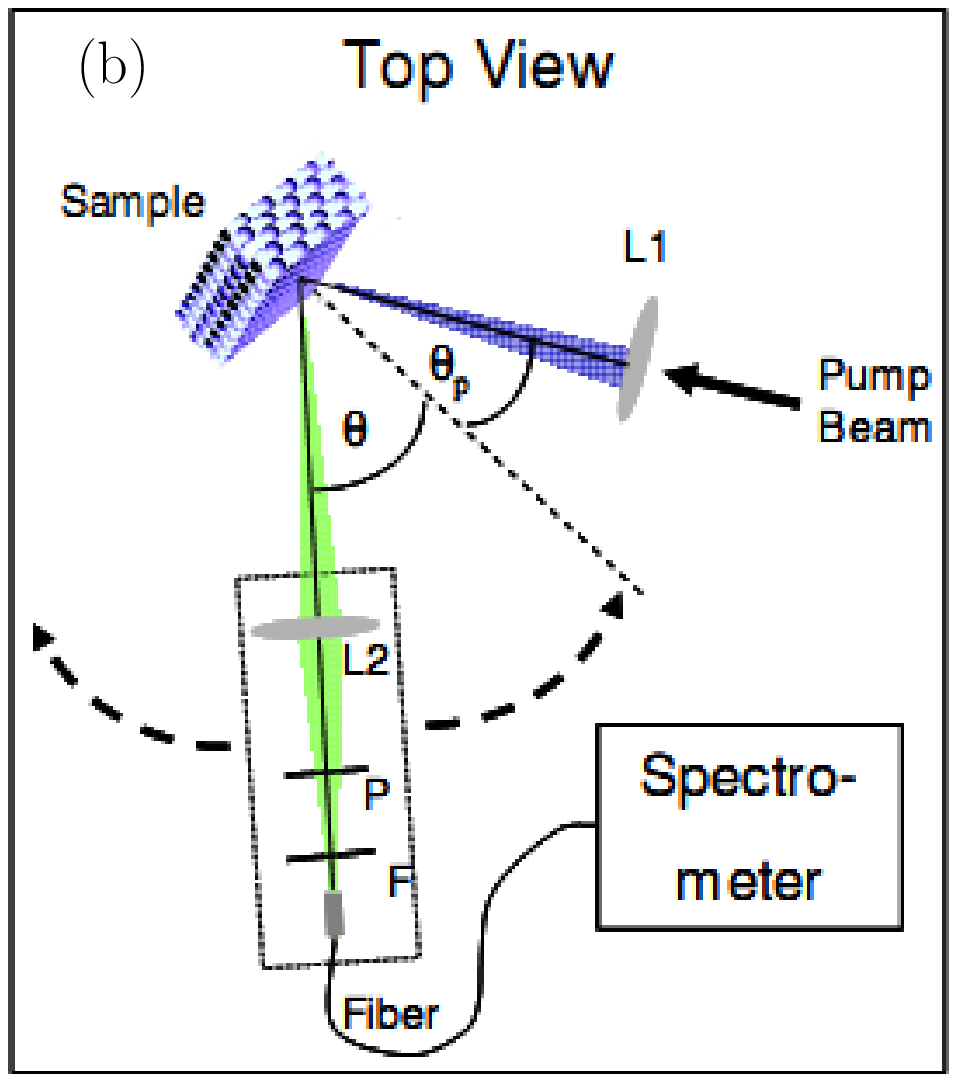}  
\end{tabular}
\caption{Schematic of experimental setup for angle- and polarization-resolved reflection (a) and PL (b) measurements.}
\label{fig:setup}
\end{figure}

We performed the angle- and polarization-resolved reflection measurements to characterize the photonic band structures. Figure \ref{fig:setup}(a) shows the experimental setup. The sample was mounted on a goniometer stage.  The reflection scan was taken in the $\theta - 2 \theta$ geometry, namely, when the sample was rotated angle $\theta$, the detection arm was rotated $2 \theta$. Collimated white light from a UV enhanced Xe lamp was focused onto the sample by a lens (L1). The beam spot on the sample surface was about 1mm in diameter. The angle of incidence from the surface normal was $\theta$. The reflected light was collected by another lens (L2) and focused to a fiber bundle which was connected to a spectrometer. A linear polarizer (P) was placed in front of the fiber bundle to select $s$- or $p$-polarized light with electric field perpendicular or parallel to the detection plane (made of the detection arm and the normal of sample surface). The angular resolution, which was determined mainly by the collection angle of lens L2, was about $5^{\circ}$. The reflection spectra were taken with the incident angle $\theta$ varying from  $5^{\circ}$ to $60^{\circ}$. The spectral resolution was about 1 nm.

In the photoluminescence (PL) experiment, only the detection arm moved and the sample did not rotate. As shown in Figure \ref{fig:setup}(b), the white light was replaced with the He:Cd laser light which excited the ZnO. The pump beam was focused onto the sample at a fixed angle $\theta_p \sim 30^{\circ}$. Spectra of emission into different angle $\theta$ were measured when the detection arm was scanned in the horizontal plane. To prevent the reflected pump light from entering the detector, the incident beam is shifted vertically so that the incidence plane deviates from the detection plane. A long pass filter (F) was placed in front of the fiber bundle to block the scattered pump light at wavelength $\lambda = 325$ nm. 

\subsection{Reflection spectra}

\begin{figure}[htbp]
\begin{tabular}{c c}
\includegraphics[width=80mm]{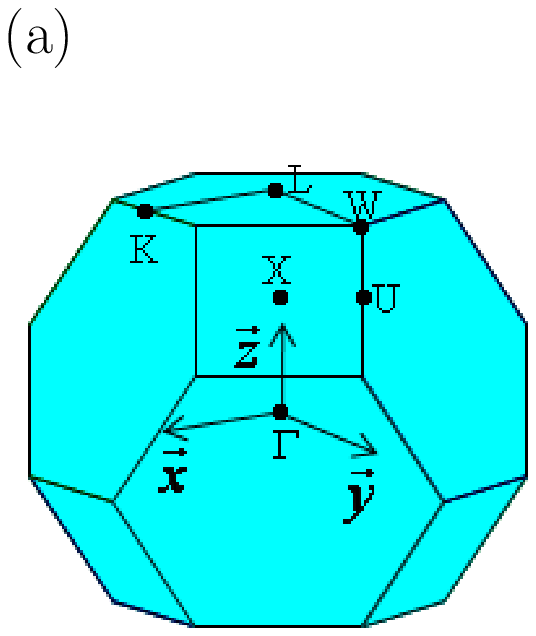} & 
\includegraphics[width=80mm]{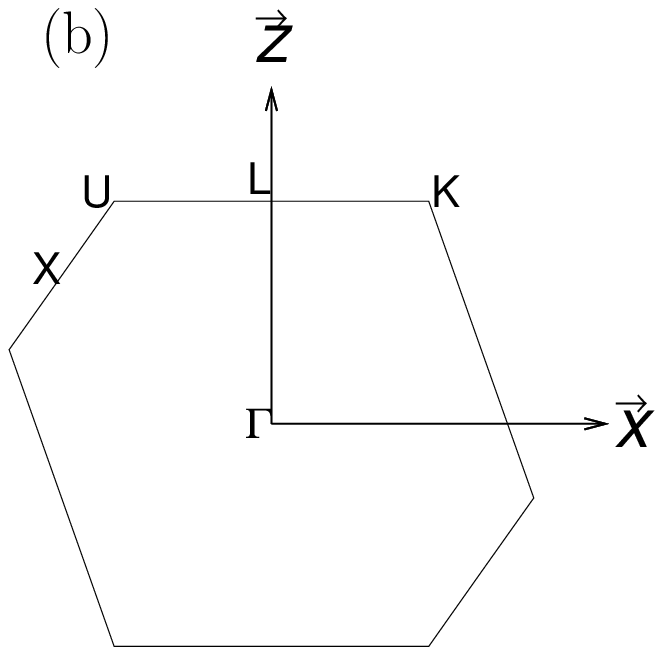} 
\end{tabular}
\caption{(a) The first Brillouin zone (BZ) for a FCC structure. The symmetry points are marked.  (b) Cross section of the first BZ by the $\Gamma L K$ plane.}
\label{fig:FBZ}
\end{figure}

We measured many ZnO inverse opals with different lattice constants. The sample uniformity was checked carefully with scanning electron microscopy and optical spectroscopy. Only the samples that were uniform in crystalline arrangement, thickness, and infiltration over an area much larger than the probe beam spot were used in the reflection and PL measurements. The data presented below were taken from one sample with the air sphere diameter 400 nm and the lattice constant 566 nm. The number of layers of air spheres was about 60. Figure \ref{fig:FBZ}(a) plots the first Brillouin zone (BZ) of ZnO inverse opal. In the reflection and PL measurements, the sample was oriented so that the detector was scanned in the $\Gamma L K$ plane. Note that in the reflection measurement the UV light from the Xe lamp could excite ZnO and generate PL. The PL intensity, however, was much weaker than the reflected light intensity. Thus, the PL signals can be ignored in the reflection spectra.

\begin{figure}[htbp]
\begin{tabular}{c c}
\includegraphics[width=60mm]{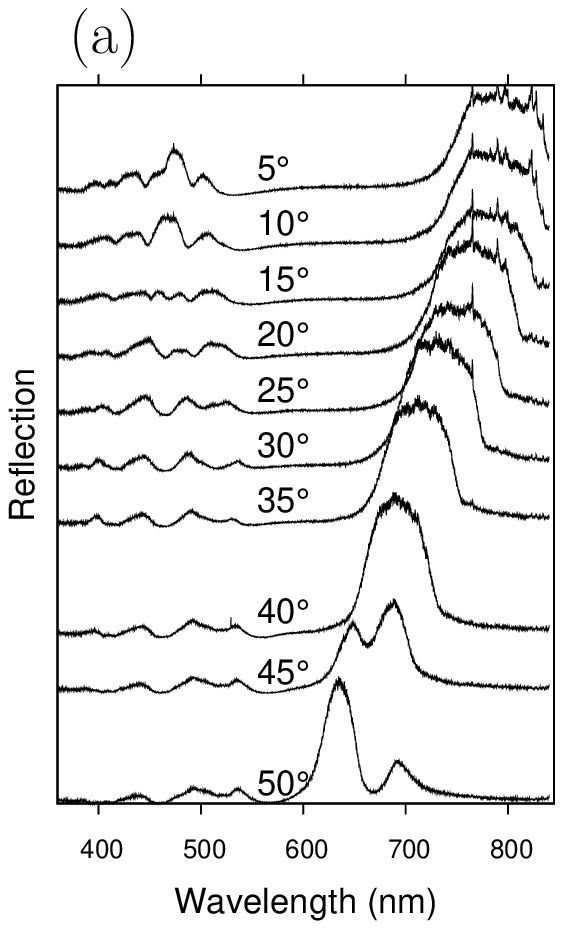}   &
\includegraphics[width=60mm]{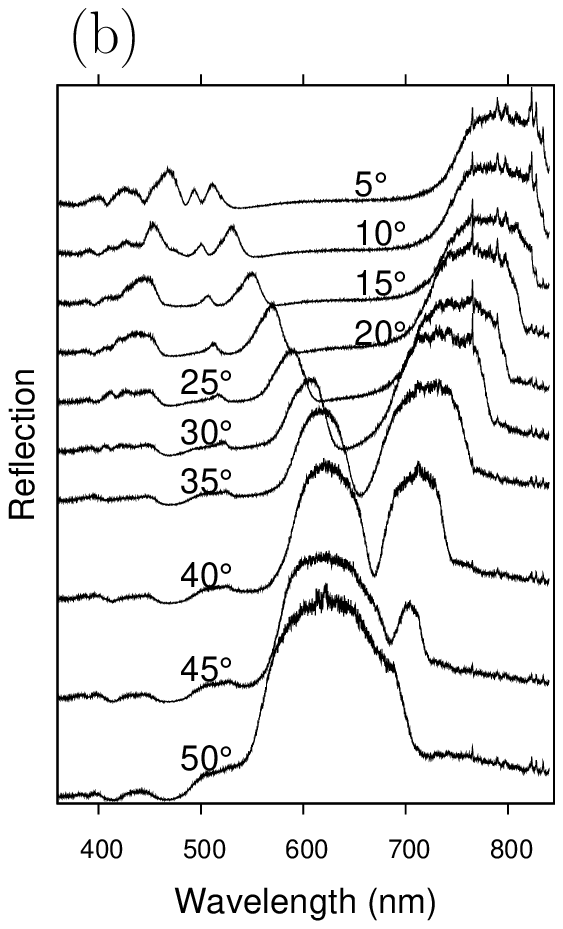}  
\end{tabular}
\caption{Angle-resolved reflection spectra of a ZnO inverse opal with sphere diameter 400 nm. Both the incidence plane and detection plane are parallel to the $\Gamma L K$ plane.  The incidence angle $\theta$ ranges from $5^{\circ}$ to $50^{\circ}$. The values of $\theta$ are written in the graph. The spectra are offset vertically. (a) $p$-polarized light, (b) $s$-polarized light. }
\label{fig:reflection}
\end{figure}

Figure \ref{fig:reflection}(a) shows the reflection spectra of $p$-polarized light for $\theta$ varying from $5^{\circ}$ to $55^{\circ}$ with $5^{\circ}$ steps. For $\theta = 5^{\circ}$, there is a primary reflection peak centered at wavelength $\lambda \simeq 800$ nm with the peak reflectivity $\sim$ 0.7. It corresponds to the lowest-order photonic band gap in [111] direction, which is confirmed in the numerical simulation to be presented in the next section. With increasing angle $\theta$, this peak shifts to shorter wavelength. There are additional reflection peaks at higher frequencies. They are less dispersive and shift to slightly longer wavelengths with increasing $\theta$. The reflection spectra of $s$-polarized light, shown in Fig. \ref{fig:reflection}(b), exhibited significant difference from those for $p$-polarized light. There is an additional reflection peak at $\lambda \simeq 510$ nm for $\theta = 5^{\circ}$. It shifts significantly to longer wavelengths with increasing $\theta$. Since the primary reflection peak shifts to the opposite direction, these two peaks exhibit an anti-crossing at $\theta \sim 45^{\circ}$. 

\subsection{Photoluminescence data}

When pumped by the He:Cd laser at $\lambda = 325$ nm, the ZnO inverse opal has PL in both the UV and visible frequencies. The UV emission is ascribed to electron transition from the conduction band to valence band, and the visible emission is via the defect states within the electronic bandgap. \cite{ozgur_comprehensive_2005,ong_evolution_2004,lin_optical_2006,lim_photoluminescence_2004,guo_intensity_2003,djurisic_defect_2007,djurisic_optical_2006} At room temperature the defect emission, which results from various material defects such as oxygen vacancies, zinc interstitials and oxygen interstitials, covers a 300 nm wavelength range. In this paper we concentrate on the defect emission, whose broad spectral range allows us to observe the effects of many different-order photonic bands on emission.  

\begin{figure}[htbp]
\includegraphics[width=100mm]{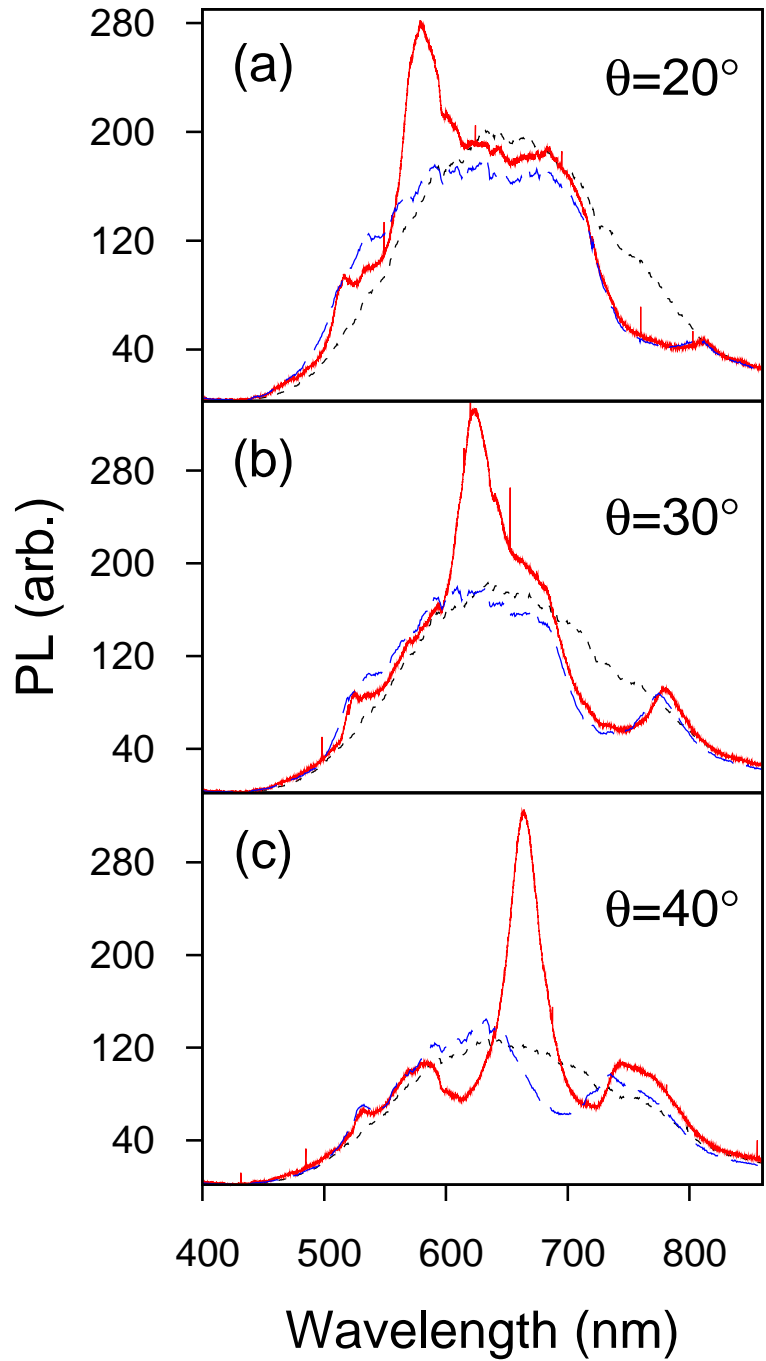}
\caption{(Color online) Measured PL spectra of a ZnO inverse opal (sphere diameter = 400 nm) and a random sample (black dotted line). The emission angle $\theta = 20^{\circ}$ (a), $30^{\circ}$ (b), and $40^{\circ}$ (c). The red solid curve and blue dashed curve represent the $s$ and $p$-polarized emission from the ZnO inverse opal, respectively. }
\label{fig:PL}
\end{figure}

Figure \ref{fig:PL} shows the PL spectra of the ZnO inverse opal and the random sample. The spectral shape of PL from the random sample does not change with observation angle $\theta$. The PL of ZnO inverse opal is strongly modified, and the modification is angle-dependent. For comparison, the emission spectra taken at identical $\theta$ are scaled so that they overlap at $\lambda = 880$ nm, well below the lowest-order PBG. Suppression of emission at longer wavelength is evident in the ZnO inverse opal. At shorter wavelength there is significant enhancement of s-polarized emission.  This enhancement is not related to stimulated emission, as the emission intensity is confirmed to vary linearly with pump intensity.    

\begin{figure}[htbp]
\begin{tabular}{c c}
\includegraphics[width=59mm]{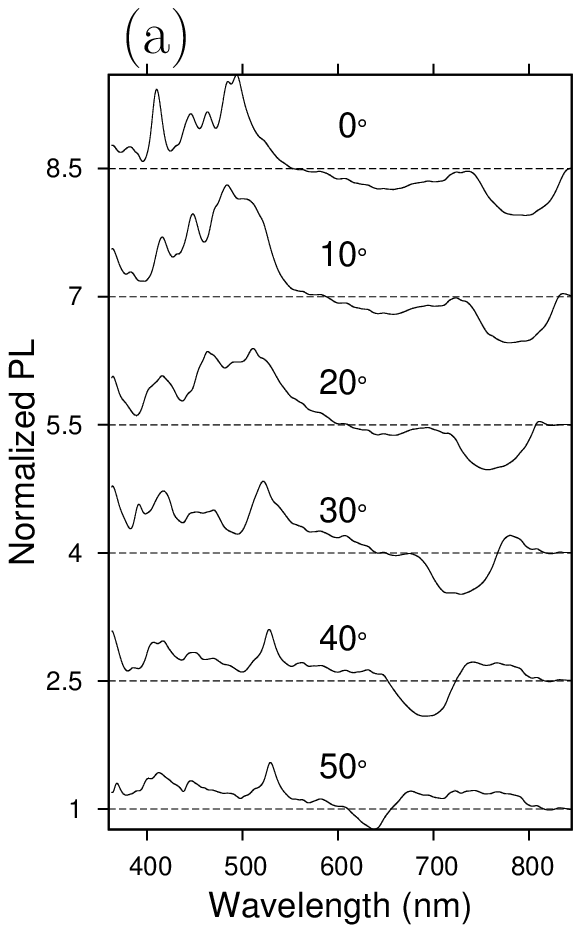} &
\includegraphics[width=60mm]{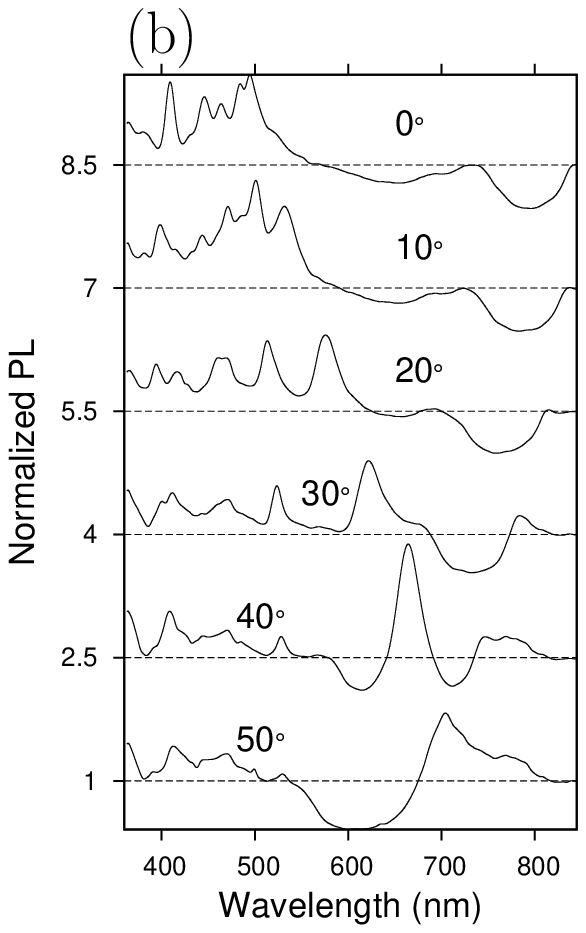}
\end{tabular}
\caption{Normalized PL spectra of the ZnO inverse opal at various emission angles $\theta$. The values of $\theta$ are written in the graph. The spectra are vertically shifted with a constant offset of 1.5. The reference line of unity for each spectrum is plotted as a horizontal dashed line. (a) $p$-polarized emission, (b) $s$-polarized emission.}
\label{fig:PL1}
\end{figure}

We extracted the PL enhancement factor by dividing the PL spectrum of ZnO inverse opal by that of the reference sample for same $\theta$. If the normalized PL intensity is less (or more) than unity, the spontaneous emission is suppressed (or enhanced). Figure \ref{fig:PL1} shows the normalized PL spectra for both $s$- and $p$-polarizations with $\theta$ varying from $0^{\circ}$ to $50^{\circ}$. The spectra are vertically shifted with a constant offset of 1.5. The reference line of unity for each spectrum is plotted as a dashed line. For $\theta = 10^{\circ}$, the dip at $\lambda \sim 800$ nm coincides with the primary reflection peak in Fig. \ref{fig:reflection}. Its blue-shift with increasing  $\theta$ is identical to that of the reflection peak. As mentioned earlier, the primary reflection peak corresponds to the lowest-order PBG. This partial gap suppresses the emission due to depletion of density of states (DOS) within certain angle range. The peaks at higher frequencies in the normalized PL spectra reveal the emission enhancement by higher-order photonic band structures. Most enhancement peaks for both polarizations are weakly dispersive with angle, except one for  $s$-polarized PL. This peak red-shifts dramatically with increasing $\theta$. Its amplitude reaches a maximal value of 2.3 at $\theta = 40^{\circ}$, exceeding all other peaks. It is responsible for strong enhancement of s-polarized PL in Fig. \ref{fig:PL}. We notice that this PL peak has similar dispersion to the reflection peak that exists only for $s$-polarization in  Fig. \ref{fig:reflection}. To compare their frequencies, we overlay the normalized PL spectra and reflection spectra in Fig. \ref{fig:GLK_s} (1st row) for $\theta = 20^{\circ} - 50^{\circ}$. It is evident that the two peaks do not overlap spectrally, instead the PL peak is always at the low frequency shoulder of the reflection peak. 

\section{Numerical simulation}

\subsection{Reduced photonic bands}

To interpret the experimental results, we performed numerical simulations of ZnO inverse opals using the experimental values. To account for the angular dependence of reflection and PL, we calculated photonic bands and density of states (DOS) corresponding to a specified angle of incidence/exit in air. These bands and DOS are called reduced bands and reduced DOS \cite{pavarini_band_2005}, as opposed to angle-integrated bands and DOS. Since the top surface of ZnO inverse opal corresponds to (111) crystallographic plane, we define a reference frame shown in Fig. \ref{fig:FBZ}: the origin at the $\Gamma$ point, $z$-axis in the $\Gamma L$ direction normal to (1 1 1) surface, $xy$ plane parallel to the surface with $x$-axis in the $L K$ direction and $y$-axis in the $L W$ direction. The wavevector of an incident/exit photon in air can be expressed as 
\begin{equation}
{\bf k}_{i} = { \frac{\omega}{c}} \left(\sin \theta \cos \phi, \sin \theta \sin \phi, \cos \phi \right),
\end{equation}
where $\omega$ is the frequency, $\theta$ is the polar angle, $\phi$ is the azimuthal angle. It should be noted that $\theta$ and $\phi$ denote the direction of light incidence/exit in air, which differs from the direction of light propagation inside the PhC. Even if the  incidence/exit direction in air is fixed, the propagation direction inside the PhC may change with frequency $\omega$.   


The reduced photonic band structure is obtained from the full band structure by using two conservation laws: (i) parallel momentum conservation modulo a reciprocal lattice vector, (ii) energy conservation. They can be written as
\begin{equation}
\textbf{k}_{\Vert}+\textbf{G}_{\Vert}=\frac{\omega}{c} \left( \sin \theta \cos \phi, \sin \theta \sin \phi, 0\right) \, ,
\end{equation}
and
\begin{equation}
\epsilon_{n}(\textbf{k}_{\Vert},\textbf{k}_{z})=h\omega \, ,
\end{equation}
where $\textbf{k}_{\Vert}$ is the \textbf{k}-component parallel to the (111) surface ($xy$-plane),  $\textbf{G}_{\Vert}$ is the parallel component of any reciprocal lattice vector $\textbf{G}$, and $\epsilon_{n}(\textbf{k}_{\Vert},\textbf{k}_{z})$ represents the $n$th energy band of  PhC with wavevector $\textbf{k}= \textbf{k}_{\Vert} + \textbf{k}_{z}$. 

We used the computer program MULTEM, which is based on the layer KKR (Korringa-Kohn-Rostoker) method \cite{stefanou_multem_2000}, to calculate the reduced band structure. One advantage of this method is that it gives $\textbf{k}_{z}$ for given $\epsilon_{n}$ and $\textbf{k}_{\Vert}$. Another advantage is that it can calculate angle-resolved reflection and transmission spectra of a PhC with finite thickness, thus allowing direct comparison with the experimental data. The reduced DOS is inversely proportional to the group velocity, which is calculated from  the slope of dispersion curve for individual reduced band. 
Since the frequency range of calculation is below the ZnO electronic bandgap, the dispersion of ZnO refractive index is rather weak. We neglected the frequency dependence of refractive index in the numerical simulation. To take into account the interstitial tetrahedral pores in the ZnO inverse opals \cite{scharrer_fabrication_2005}, we set the value of refractive index  at 1.95, which is slightly lower than the actual refractive index of ZnO.        

\begin{figure}[htbp]
\begin{center}
\includegraphics[width=150mm]{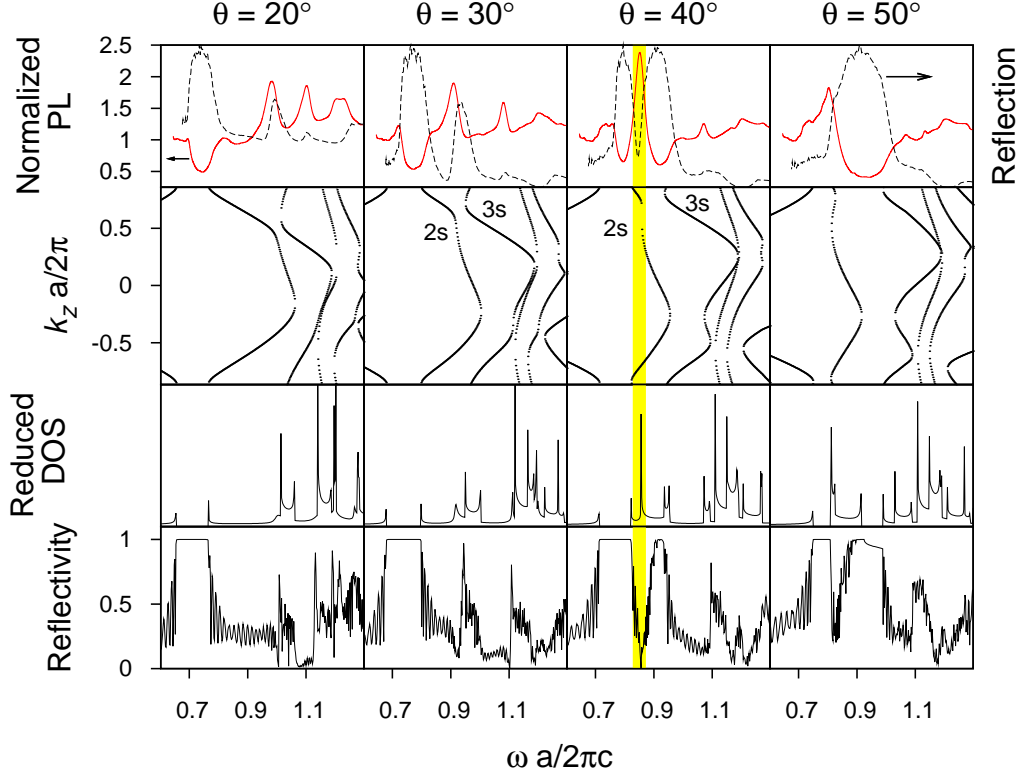} 
\end{center}
\caption{(Color online) 1st row: angle-resolved $s$-polarized reflection spectra (black dashed line) of the ZnO inverse opal overlaid with the normalized PL spectra (red solid line) of same polarization and  angle $\theta$. 2nd row: calculated $s$-polarized reduced band structure of the ZnO inverse opal for a fixed angle of incidence/exit in air. Lattice constant $a= 566$ nm, the dielectric constant of ZnO is $\varepsilon = 3.8$. 3rd row: calculated reduced density of $s$-polarized states of the ZnO inverse opal. 4th row: calculated reflectivity of $s$-polarized light from a ZnO inverse opal whose thickness is 34 layers of air spheres.  $\phi = 0$, and $\theta = 20^{\circ}$ (1st column), $30^{\circ}$ (2nd column), $40^{\circ}$ (3rd column), $50^{\circ}$ (4th column). For $\theta= 40^\circ$, a stationary inflection point is developed for the $2s$ band at $\omega a / 2 \pi c = 0.856$.}
\label{fig:GLK_s}
\end{figure}

\begin{figure}[htbp]
\begin{center}
\includegraphics[width=150mm]{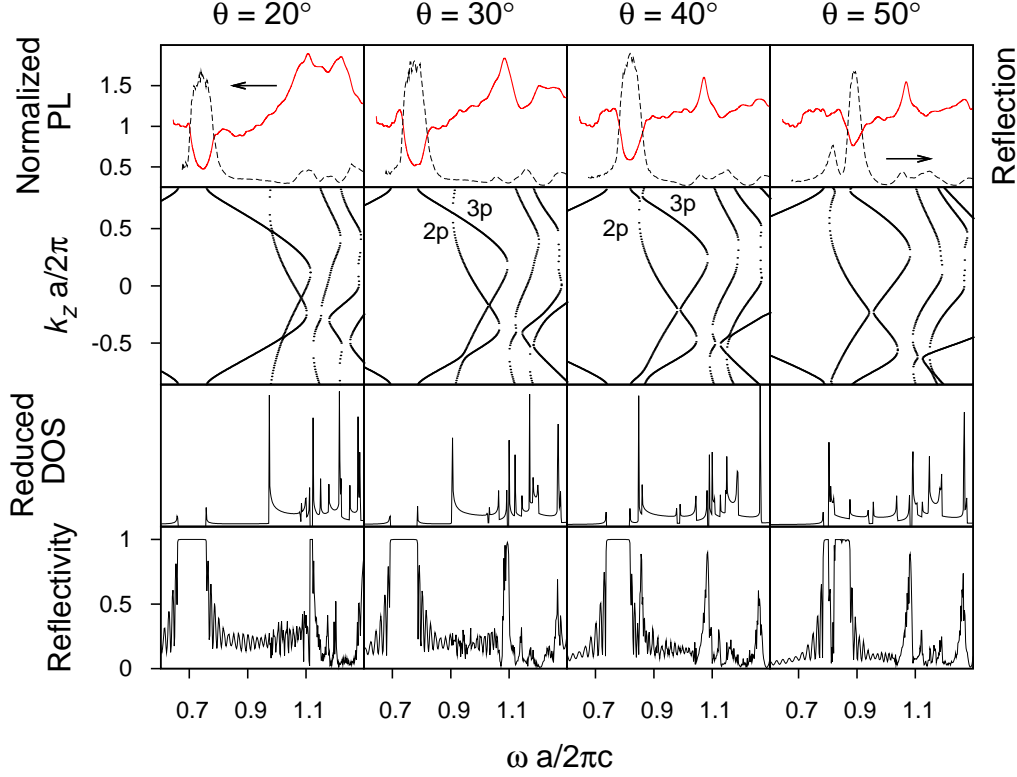} \\
\end{center}
\caption{Same as Fig. \ref{fig:GLK_s} for $p$-polarization. }
\label{fig:GLK_p}
\end{figure}

We calculated the reduced photonic bands for ${\bf k}_{i}$ scanning in the $\Gamma L K$ plane, namely $\phi = 0$ and $\theta$ varying from $0^{\circ}$ to $80^{\circ}$ with $10^{\circ}$ steps. The first BZ has mirror symmetry with respect to the $\Gamma L K$ plane [Fig.  \ref{fig:FBZ}(a)]. The $s$- or $p$-polarized light, with ${\bf k}_i$ in the $\Gamma L K$ plane and electric field perpendicular or parallel to the $\Gamma L K$ plane, has distinct mirror symmetry with respect to the $\Gamma L K$ plane. Hence, the two polarizations are decoupled and the reduced photonic bands with ${\bf k}_i$ in the $\Gamma L K$ plane are either $s$- or $p$-polarized. We plot the $s$-polarized bands in Fig. \ref{fig:GLK_s} (2nd row)  and $p$-polarized bands in Fig. \ref{fig:GLK_p} (2nd row). The reduced bands with positive $k_z$ differ from those with negative $k_z$. This is because the cross section of the first BZ by the $\Gamma L K$ plane, shown in Fig. \ref{fig:FBZ}(b), does not have mirror symmetry with respect to $x$-axis. 

Comparing the reduced $s$- and $p$-polarized bands reveals their significant differences. For example, at $\theta = 30^{\circ}$, the second and third $s$-polarized bands [labeled as $2s$ and $3s$ in Fig. \ref{fig:GLK_s}] exhibit frequency anti-crossing around $k_z a/ 2 \pi = 0.7$, while $2p$ and $3p$ bands nearly cross in Fig. \ref{fig:GLK_p}. For both polarizations, the second and third reduced bands originate from Bragg diffraction of light  by (1 1 1) and (${\rm \overline{1}}$ 1 1) planes for $k_z>0$, and by (1 1 1) and (2 0 0) planes for $k_z < 0$.  \cite{baryshev_polarized_2007} At their crossing point near $k_z a / 2 \pi \sim 0.7$, simultaneous Bragg diffraction by (1 1 1) and (${\rm \overline{1}}$ 1 1) planes results in band repulsion. \cite{van_driel_multiple_2000} The anti-crossing of $2s$ and $3s$ bands indicates strong band coupling via multiple diffraction. The interaction of $2p$ and $3p$ bands, however, is much weaker. Such difference can be explained by the dependence of diffraction efficiency on polarization. It is well known for $X$-ray diffraction that when the Bragg angle is close to $45^{\circ}$ the intensity of diffracted beam is extremely weak for $p$-polarized wave. \cite{morelhao_strength_2001} The suppression of Bragg diffraction has formal analogy to the Brewster effect on reflection of $p$-polarized light by homogeneous medium. The diffraction efficiency for $p$-polarized light in a PhC can be greatly reduced if the incident angle approaches the critical angle. \cite{baryshev_interaction_2006,dukin_polarization_2006} The weak coupling of $2p$ and $3p$ bands is attributed to low efficiency of multiple diffraction of $p$-polarized light at the band crossing point because one of the Bragg angles is close to the critical angle. 

Comparison of the calculated band structure to the measured reflection spectra confirms that the primary reflection peak corresponds to the lowest-order gap between the first and second bands for both polarizations. The additional reflection peak, observed only for $s$-polarization, overlaps with the gap opened by anti-crossing of $2s$ and $3s$ bands. As $\theta$ increases, the gap moves toward lower frequency, and the reflection peak follows. This reflection peak is not observed for $p$-polarization because $2p$ and $3p$ bands have little repulsion. In fact, the calculated reflection spectrum for $\theta = 40^{\circ}$ exhibits a very narrow peak corresponding to the small gap between $2p$ and $3p$ bands at $k_z a/ 2 \pi \sim 0.7$ [Fig. \ref{fig:GLK_p}]. Such narrow peak is smeared out experimentally by averaging over finite collection angle. 

\subsection{Stationary inflection point}

The dispersion of the $2s$ band is nearly flat in the vicinity of its avoided crossing with the $3s$ band. It produces a peak in the reduced DOS shown in Fig. \ref{fig:GLK_s}. The enhanced PL peak, which is observed only for $s$-polarization, coincides with this DOS peak. It follows the DOS peak as it moves to lower frequency at higher $\theta$. At $\theta = 40^{\circ}$, the dispersion curve for $2s$ band has a stationary inflection point at $k_z a / 2  \pi \simeq 0.6$ where $d\omega / dk_z \simeq 0$ and $d^{2}\omega / dk_z^{2} \simeq 0$.

The existence of stationary inflection point is verified by tracing the evolution of the $2s$ band with $\theta$. For $\theta =  50^{\circ}$, the dispersion curve for the $2s$ band exhibits a local minima at $k_z a / 2 \pi \sim 0.5$ and a local maxima at $k_z a / 2 \pi \sim 0.7$. As $\theta$ decreases, the local minima and local maxima approach each other, eventually they merge at $\theta \simeq 40^{\circ}$. With a further decrease of $\theta$, e.g. at $\theta = 30^{\circ}$, the dispersion curve has neither local minima nor local maxima, its slope does not change sign throughout the region of interest. Such band evolution confirms that the $2s$ band has a stationary inflection point at $\theta = 40^{\circ}$ where the merging of a local minima and a local maxima gives not only $d\omega / dk_z=0$ but also $d^{2}\omega/dk_z^{2}=0$. The evolution of the $2p$ band with $\theta$, shown in the 2nd row of Fig. \ref{fig:GLK_p}, is completely different from that of the $2s$ band. It reveals that the $2p$ band does not have a stationary inflection point near $\theta= 40^{\circ}$, possibly due to its tiny anti-crossing with the $3p$ band. 

The DOS at the stationary inflection point diverges in an infinite large PhC. In a real sample such divergence is avoided because of finite sample size. Nevertheless, the DOS peak has the maximal amplitude at $\theta \simeq 40^{\circ}$ where the stationary inflection point is developed. The large DOS enhances the spontaneous emission process. Although the $2p$ band does not have a stationary inflection point, its dispersion is relatively flat in the neighborhood of $k_z a / 2 \pi = 0.6$ [Fig. \ref{fig:GLK_p}]. It produces a peak in the reduced DOS, which should enhance emission. Experimentally, the $p$-polarized PL is not enhanced.  

\begin{figure}[htbp]
\begin{tabular}{c c}
\includegraphics[width=70mm]{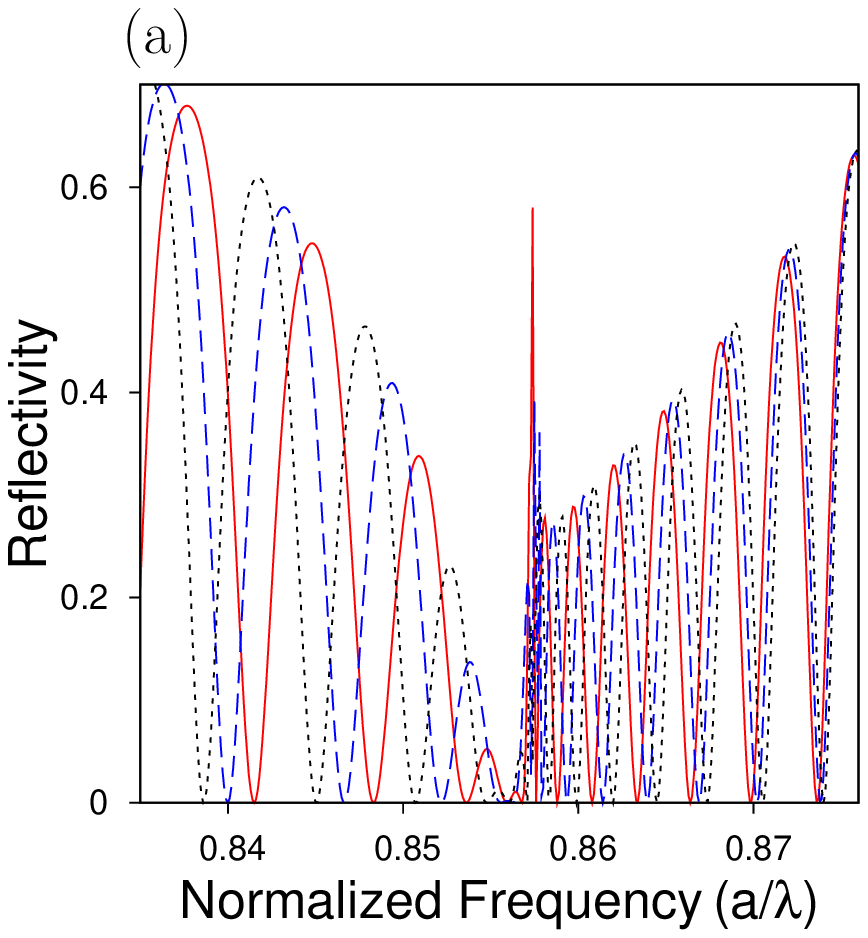} &
\includegraphics[width=70mm]{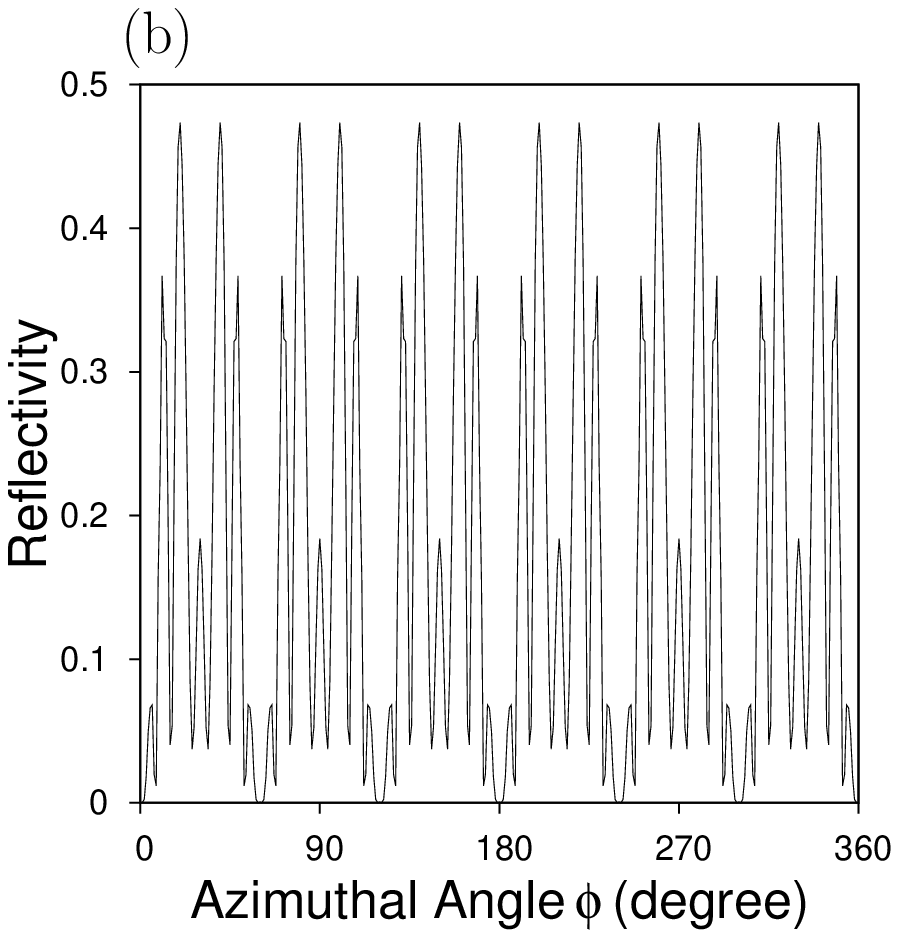}  \\
\end{tabular}
\caption{(Color online) (a) Calculated reflectivity of $s$-polarized light as a function of normalized frequency $\omega a / 2 \pi c$. The incidence angle $\theta = 40^{\circ}$, $\phi = 0$. The thickness of ZnO inverse opal is equal to 34 (red solid line), 36 (blue dashed line) and 38 (black dotted line) layers of air spheres. (b) Calculated reflectivity of $s$-polarized light at the stationary inflection point $\omega a / 2 \pi c = 0.856$ as a function of the azimuthal angle $\phi$. $\theta$ is fixed at $40^{\circ}$.}
\label{fig:layer}
\end{figure}

The question arises why the DOS peak leads to enhanced PL for $s$-polarization but not for $p$-polarization. The answer lies in the emission extraction efficiency. One unique property of the frozen mode at the stationary inflection point is its efficient coupling to the free photon mode outside the PhC. It leads to vanishing reflectivity at the sample/air interface, which is confirmed by our calculation and measurement of reflectivity from ZnO inverse opal. The calculated reflection spectrum for $\theta = 40^{\circ}$ [Fig. \ref{fig:GLK_s}] reveals that the reflectivity at the stationary inflection point $\omega a / 2 \pi c = 0.856$ is almost zero. Experimentally, the measured reflectivity is low but not zero due to averaging over finite collection angle. To verify that the zero reflectivity is not a result of finite sample thickness, we calculated the reflection spectra of three ZnO inverse opals with different thicknesses (34, 36 and 38 layers of air spheres). Figure \ref{fig:layer}(a) shows that for  $\theta = 40^{\circ}$ the reflectivity reaches zero at multiple frequencies. As the sample thickness varies, all the zero points of reflectivity shift in frequency except the one at $\omega a / 2 \pi c = 0.856$. Their dependence on sample thickness suggests those vanishing reflectivities result from interference of light multiply reflected by the two surfaces of the ZnO inverse opal. The fact that the zero reflectivity at $\omega a / 2 \pi c = 0.856$ is independent of sample thickness confirms that it is caused not by the Fabry-Perot resonance in a finite PhC slab but by the intrinsic property of photonic band, more specifically, the dispersion of $2s$ band. This result demonstrates a perfect impedance match for the frozen mode at the interface of ZnO inverse opal and air. The light emitted to the frozen mode experiences little reflection at the sample surface when leaving the sample. Hence, $s$-polarized PL at the stationary inflection point of $2s$ band can be efficiently extracted from the sample. The reflection of $p$-polarized light does not go to zero in the absence of frozen mode, thus $p$-polarized PL cannot escape easily from the sample.          

Therefore, the observed large enhancement peak of $s$-polarized PL around $\theta = 40^{\circ}$ is attributed to two factors: (i) enhanced emission into the frozen mode due to large DOS, (ii) efficient extraction of emitted light out of the PhC. As $\theta$ approaches $40^{\circ}$, the DOS increases and surface reflection decreases. Thus, PL is enhanced in the vicinity of a stationary inflection point. The maximal PL enhancement occurs at the stationary inflection point where the DOS reaches the maximum and the surface reflection the minimum. 

The enhanced emission is directional, namely, it exits the ZnO inverse opal to air at the polar angle $\theta= 40^{\circ}$. Next we investigate the emission directionality in terms of azimuthal angle $\phi$. Since the cross section of the first BZ by the $\Gamma L K$ plane does not have mirror symmetry with respect to the $z$-axis ($\Gamma L$ direction), the reduced band structure for ${\bf k}_i$ scanning along the $LK$ path is different from that along the $LU$ path. The reflection spectra, however, are the same for the two scanning directions due to reciprocity of reflection. \cite{gippius_optical_2005} This is confirmed by our calculation and measurement of reflection spectra. Figure \ref{fig:layer}(b) plots the calculated reflectivity for $s$-polarized light at the frequency of stationary inflection point versus the azimuthal angle $\phi$. The reflectivity exhibits six-fold symmetry when $\phi$ varies from $0^{\circ}$ to $360^{\circ}$. In principle, the PL spectra should exhibit difference between the scanning path along $L K$ and that along $L U$. Experimentally, the PL spectra for the two scanning paths are similar. This is attributed to the structure disorder, in particular, the twin structure in the ZnO inverse opal. \cite{baryshev_polarized_2007} As a result, the enhanced PL can be observed at six azimuthal angles.       



\section{PL enhancement in a different crystal direction}

\begin{figure}[htbp]
\begin{tabular}{c c}
\includegraphics[width=59mm]{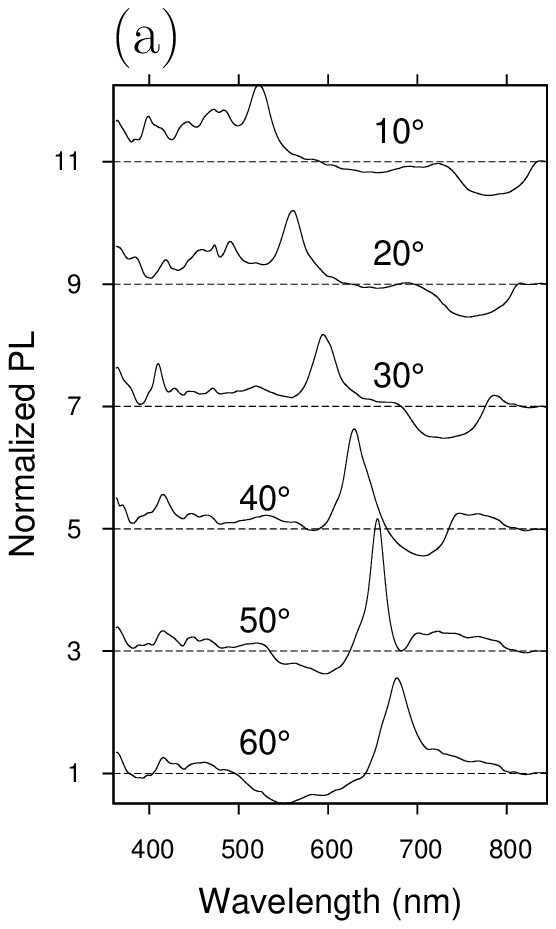}   &
\includegraphics[width=60mm]{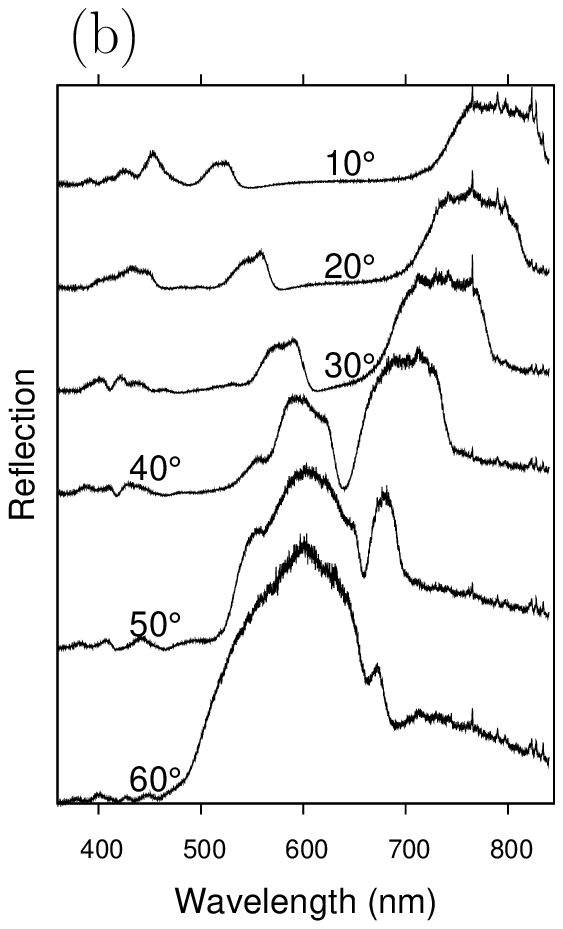}  \\
\end{tabular}
\caption{Measured angle-resolved spectra of normalized PL (a) and reflection (b) of a ZnO inverse opal with sphere diameter = 400 nm. The scanning plane is parallel to the $\Gamma LW$ plane, $\phi = 90^{\circ}$. $\theta$ varies from $10^{\circ}$ to $60^{\circ}$. The values of $\theta$ are written on the graph. In (a), the normalized PL spectra are shifted vertically with a constant offset of 2. The horizontal dashed lines mark unity for individual spectra. The reflection spectra in (b) are also shifted vertically.}
\label{fig:GLW}
\end{figure}

Although the data presented above were taken from the ZnO inverse opal with sphere diameter 400 nm, we repeated the experiments with several samples of different sphere size and obtained similar results. The enhancement of PL by the frozen mode is a common phenomenon because many high-order bands of ZnO inverse opals have stationary inflection points in their dispersion curves. In addition to scanning in the $\Gamma L K$ plane, we also scanned in the $\Gamma L W$ plane and observed PL enhancement at a different angle $\theta$. 

\begin{figure}[htbp]
\begin{center}
\includegraphics[width=150mm]{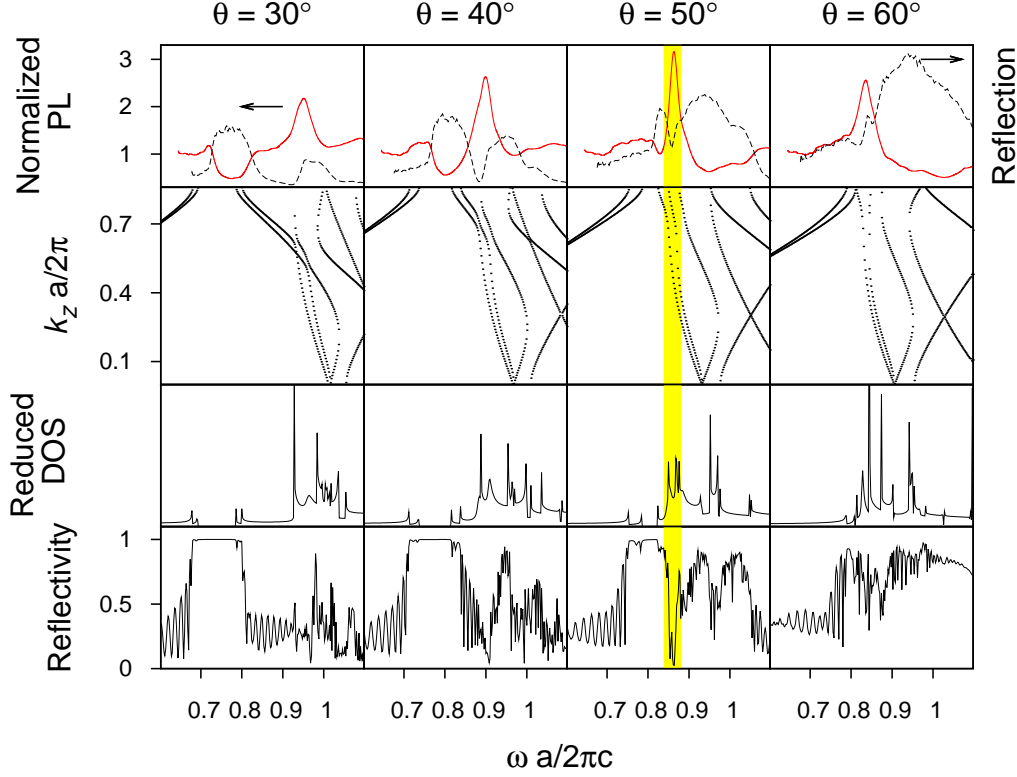}
\end{center}
\caption{(Color online) 1st row: angle-resolved reflection spectra (black dashed line) of the ZnO inverse opal overlaid with the normalized PL spectra (red solid line) of same angle $\theta$. 2nd row: calculated reduced band structure of the ZnO inverse opal for a fixed angle of incidence/exit in air. Lattice constant $a= 566$ nm, the dielectric constant of ZnO $\varepsilon = 3.8$. 3rd row: calculated reduced density of states of the ZnO inverse opal. 4th row: calculated reflectivity of a ZnO inverse opal with thickness = 34 layers of air spheres.   $\phi = 90^{\circ}$, and $\theta = 30^{\circ}$ (1st column), $40^{\circ}$ (2nd column), $50^{\circ}$ (3rd column), $60^{\circ}$ (4th column). For $\theta= 50^\circ$, a stationary inflection point is developed at $\omega a / 2 \pi c = 0.87$.}
\label{fig:GLW_all}
\end{figure}

Fig. \ref{fig:GLW}(a) shows the angle-resolved PL spectra when scanning in the $\Gamma L W$ plane. The normalized PL spectra are shifted vertically with a constant offset of 2. We observed an enhancement peak which shifts to lower frequency with increasing $\theta$. The maximum enhancement occurred around $\theta = 50^{\circ}$ and reached the value of 3.2. Since the cross section of the first BZ by the $\Gamma L W$ plane is symmetric with respect to $y$-axis (parallel to $\Gamma$L direction), the photonic mode with $({\bf k}_{||}, k_z)$ is identical to that with $({\bf k}_{||}, -k_z)$. The reduced band structure with positive $k_z$ is the same as that with negative $k_z$. However, the first BZ does not have mirror symmetry with respect to the $\Gamma L W$ plane. The photonic modes with ${\bf k}_i$ in the $\Gamma L W$ plane contain both electric field components parallel and perpendicular to the $\Gamma L W$ plane. Hence, the reduced photonic bands can no longer be divided into $s$- and $p$-polarized bands. Consequently, the PL spectra as well as the reflection spectra are insensitive to polarization. The angle-resolved reflection spectra in Fig. \ref{fig:GLW}(b) exhibit the frequency shift of the first and second reflection peaks with $\theta$.  Comparing the  reflection spectra to the PL spectra in Fig. \ref{fig:GLW_all} reveals that the enhanced PL peak overlaps with the reflection dip at the low-frequency side of the second reflection peak. Our calculation of the reduced band structure, the reduced DOS and reflection spectra (Fig. \ref{fig:GLW_all}) illustrates that the maximal PL enhancement peak observed around $\theta = 50^{\circ}$ results from the stationary inflection point of a high-order band, which produces a peak in DOS and a dip in reflectivity at $\omega a / 2 \pi c \simeq 0.87$. Note that the reflectivity at $\theta = 50^{\circ}$ is not zero, because the exact angle for the stationary inflection point is $47^{\circ}$. The comparison of numerical results to experimental data confirm that the PL enhancement originates from the frozen modes, similar to that observed when scanning in the $\Gamma L K$.

\section{Conclusion}

In conclusion, we present a detailed study on the angle- and polarization-resolved  photoluminescence and reflection spectra of ZnO inverse opals. The broad ZnO defect emission exhibits multiple enhancement peaks for both polarizations. Our numerical simulation reveals that the largest enhancement peak results from the frozen mode at the stationary inflection point of dispersion curve for a high-order photonic band. The frozen mode has well defined propagation direction and may be linearly polarized. At its frequency, the reduced DOS diverges, greatly enhancing the spontaneous emission into the frozen mode. Perfect coupling of the frozen mode to the free photon mode outside the sample leads to efficient extraction of emission from the PhC. The enhanced emission not only has good directionality but also can be linearly polarized. We note that the above mechanism does not work for all the enhanced PL peaks. Many peaks, which overlap spectrally with the high-order photonic bands, remain to be explained. Nevertheless, our results demonstrate that a high-order band of a 3D PhC can strongly modify the spontaneous emission, which offers potential application for a highly efficient light source.

The authors acknowledge Prof. A. Figotin, Dr. I. Vitebskiy and Dr. M. V. Erementchouk for stimulating discussions. This work was supported by the National Science Foundation under Grant Nos. ECS-0601249 and DMR-0704962.

\bibliography{sip}

\end{document}